\documentclass{article}
\usepackage{spconf,amsmath,amssymb,graphicx,url}
\usepackage{tikz}
\usepackage{comment}
\usepackage{ascmac}
\usepackage{subfigure}
\usepackage{CJKutf8} 
\usepackage{multicol}
\usepackage{multirow}
\usepackage{colortbl}
\usepackage{bbding}
\usepackage{pifont}
\usepackage{wasysym}
\usepackage{amssymb}
\usepackage[
    backend=biber,
    style=ieee,
    maxbibnames=3,
    maxcitenames=3,
    doi=false,isbn=false,url=false,eprint=false
]{biblatex} 
\usepackage{hyperref}
\hypersetup{
    colorlinks = false,
}

\newcommand{\Fig}[1]{Figure~\ref{fig:#1}} 
\newcommand{\Table}[1]{Table~\ref{tab:#1}} 
\newcommand{\Sec}[1]{Section~\ref{sec:#1}} 

\newcommand{\drawfig}[4]{ 
    \begin{figure}[#1]
    \centering 
    \vspace{0mm}
    \includegraphics[width=#2,clip]{#3.pdf} 
    \vspace{-3.5mm}
    \caption{#4}
    \label{fig:#3}
    \vspace{-3mm}
    \end{figure}
}

\DeclareSourcemap{
	\maps[datatype=bibtex, overwrite=true]{
		\map{
			\step[fieldsource=booktitle,
			match=\regexp{.*Interspeech.*},
			replace={Proc. Interspeech}]
			\step[fieldsource=journal,
			match=\regexp{.*INTERSPEECH.*},
			replace={Proc. Interspeech}]
			\step[fieldsource=booktitle,
			match=\regexp{.*ICASSP.*},
			replace={Proc. ICASSP}]
			\step[fieldsource=booktitle,
			match=\regexp{.*icassp_inpress.*},
			replace={Proc. ICASSP (in press)}]
			\step[fieldsource=booktitle,
			match=\regexp{.*Computer.*Vision.*and.*Pattern.*Recognition.*},
			replace={Proc. CVPR}]
			\step[fieldsource=booktitle,
			match=\regexp{.*International.*Conference.*on.*Acoustics,.*Speech.*and.*Signal.*Processing.*},
			replace={Proc. ICASSP}]
                \step[fieldsource=booktitle,
			match=\regexp{.*International.*Conference.*on.*Learning.*Representations.*},
			replace={Proc. ICLR}]
			\step[fieldsource=booktitle,
			match=\regexp{.*International.*Conference.*on.*Machine.*Learning.*},
			replace={Proc. ICML}]
			\step[fieldsource=booktitle,
			match=\regexp{.*Automatic.*Speech.*Recognition.*and.*Understanding.*},
			replace={Proc. ASRU}]
			\step[fieldsource=booktitle,
			match=\regexp{.*Spoken.*Language.*Technology.*},
			replace={Proc. SLT}]
			\step[fieldsource=booktitle,
			match=\regexp{.*Speech.*Synthesis.*Workshop.*},
			replace={Proc. SSW}]
			\step[fieldsource=booktitle,
			match=\regexp{.*workshop.*on.*speech.*synthesis.*},
			replace={Proc. SSW}]
			\step[fieldsource=booktitle,
			match=\regexp{.*Advances.*in.*neural.*information.*processing.*},
			replace={Proc. NIPS}]
			\step[fieldsource=booktitle,
			match=\regexp{.*Advances.*in.*Neural.*Information.*Processing.*},
			replace={Proc. NeurIPS}]
			\step[fieldsource=journal,
			match=\regexp{.*Advances.*in.*Neural.*Information.*Processing.*},
			replace={Proc. NeurIPS}]
			\step[fieldsource=booktitle,
			match=\regexp{.*Workshop.*on.* Applications.* of.* Signal.*Processing.*to.*Audio.*and.*Acoustics.*},
			replace={Proc. WASPAA}]
			\step[fieldsource=booktitle,
			match=\regexp{.*International.*Conference.*on.*Language.*Resources.*and.*Evaluation.*},
			replace={Proc. LREC}]
			\step[fieldsource=booktitle,
			match=\regexp{.*International.*Workshop.*on.*Machine.*Learning.*for.*Signal.*Processing.*},
			replace={Proc. MLSP}]
			\step[fieldsource=booktitle,
			match=\regexp{.*European.*Conference.*on.*Computer.*Vision.*},
			replace={Proc. ECCV}]
			\step[fieldsource=booktitle,
			match=\regexp{.*Applications.*of.*Computer.*Vision.*},
			replace={Proc. WACV}]
			\step[fieldsource=booktitle,
			match=\regexp{.*Detection.*and.*Classification.*of.*Acoustic.*Scenes.*and.*Events.*},
			replace={Proc. DCASE}]
			\step[fieldsource=booktitle,
			match=\regexp{.*International.*Conference.*on.*Human-Computer.*Interaction.*},
			replace={Proc. HCII}]
			\step[fieldsource=booktitle,
			match=\regexp{.*Symposium.*on.*Interactive.*3D.*Graphics.*and.*Games.*},
			replace={Proc. I3D}]
			\step[fieldsource=publisher,
			match=\regexp{.+},
			replace={{}}]
			\step[fieldsource=month,
			match=\regexp{.+},
			replace={{}}]
			\step[fieldsource=location,
			match=\regexp{.+},
			replace={{}}]
			\step[fieldsource=address,
			match=\regexp{.+},
			replace={{}}]
			\step[fieldsource=organization,
			match=\regexp{.+},
			replace={{}}]
			\step[fieldsource=doi,
			match=\regexp{.+},
			replace={{}}]
			\step[fieldsource=url,
			match=\regexp{.+},
			replace={{}}]
			\step[fieldsource=editor,
			match=\regexp{.+},
			replace={{}}]
		}
	}
} 

\title{Visual onoma-to-wave: environmental sound synthesis\\from visual onomatopoeias and sound-source images}
\name{\vspace{-0.3mm}\begin{tabular}{c} 
    Hien Ohnaka$^{1,2}$, 
    Shinnosuke Takamichi$^2$, 
    Keisuke Imoto$^3$, \\
    Yuki Okamoto$^4$, 
    Kazuki Fujii$^2$, 
    Hiroshi Saruwatari$^2$
    \end{tabular}
    \thanks{This work was supported by JSPS KAKENHI Grant Number 21H04900 and 22H03639, and ROIS NII Open Collaborative Research 2022-22S0503.}
}
\address{
    $^1$National Institute of Technology, Tokuyama College, Japan.
    $^2$The University of Tokyo, Japan.\\
    $^3$Doshisha University, Japan.
    $^4$Ritsumeikan University, Japan.
}
\newcommand{\jp}[1]{ \begin{CJK}{UTF8}{ipxm}#1\end{CJK} }

\begin{document}
\ninept
\maketitle
\setlength{\tabcolsep}{1.1mm}

\begin{abstract}
    \vspace{-1.5mm}
    We propose a method for synthesizing environmental sounds from visually represented onomatopoeias and sound sources.
    An onomatopoeia is a word that imitates a sound structure, i.e., the text representation of sound. 
    From this perspective, onoma-to-wave has been proposed to synthesize environmental sounds from the desired onomatopoeia texts.
    Onomatopoeias have another representation: visual-text representations of sounds in comics, advertisements, and virtual reality.
    A visual onomatopoeia (visual text of onomatopoeia) contains rich information that is not present in the text, such as a long-short duration of the image, so the use of this representation is expected to synthesize diverse sounds. 
    Therefore, we propose visual onoma-to-wave for environmental sound synthesis from visual onomatopoeia.
    The method can transfer visual concepts of the visual text and sound-source image to the synthesized sound.
    We also propose a data augmentation method focusing on the repetition of onomatopoeias to enhance the performance of our method.
    An experimental evaluation shows that the methods can synthesize diverse environmental sounds from visual text and sound-source images.
    
\end{abstract} 
    

\vspace{-4mm}
\section{Introduction} \vspace{-3mm}\label{sec:introduction}
Environmental sound synthesis aims to synthesize natural and diverse environmental sounds for sound-content production such as games~\cite{lloyd2011sound}.
Among several deep learning methods for environmental sound synthesis, synthesis from sound event labels (i.e., discrete symbols representing sound events) is the basic approach~\cite{kong2019sampleRNN,liu2021csg}.
This method synthesizes a sound that appropriates the given label but cannot control the detailed structure of the sound, e.g., duration and temporal changes in tone.
Another method is synthesis from a text~\cite{yang2022diffsound,anony2023audiogen,okamoto2022Onoma}.
For synthesizing sounds, onoma-to-wave~\cite{okamoto2022Onoma} (\Fig{fig/concept_paper}~(a)) uses an onomatopoeia, a word imitating the sound (i.e., the text representation of sounds).
This method can control the detailed structure of the synthesized sound by the input onomatopoeia text.
Also, the overall impression of the sound can be controlled by conditioning the auxiliary sound event label.

\drawfig{t}{0.95\linewidth}{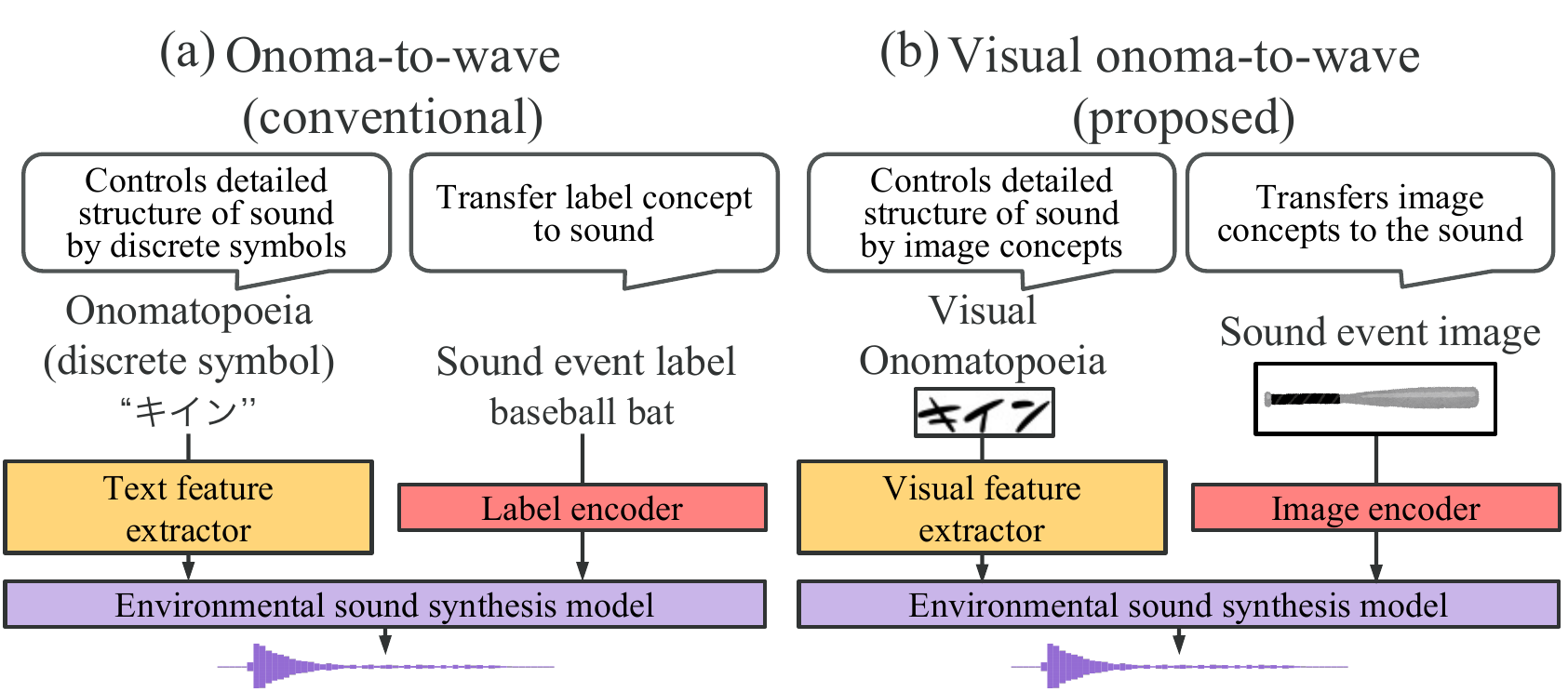}
{Method comparison. \jp{キイン} pronounced /kiin/, which is onomatopoeia of hitting baseball with metal bat.\vspace{-2mm}}

Onomatopoeias have another representation: visual-text representation of the sound.
For example, in comics and mangas, visual onomatopoeias are drawn with arbitrary shapes, sizes, and fonts for delivering sounds to the reader~\cite{baek2022COO}.
A visual onomatopoeia enables dramatic expression and draws the reader in a way that descriptions of non-textual images (e.g., sounding objects or people) alone cannot.
Visual concepts, e.g., size and text stretch, of a visual onomatopoeia evokes in the reader sound concepts, e.g., loudness and duration~\cite{huang2014fluidity}.
Another example is in virtual reality (VR).
Presenting visual onomatopoeias along with environmental sounds in immersive VR can enhance the immersive qualities of VR~\cite{oh2019VReffect}.
These examples show the possibility of applying environmental sound synthesis from arbitrary visual onomatopoeias to audible comics~\cite{wang2019comic} and immersive VR and motivate us to develop a synthesis method. Also, since visual onomatopoeias frequently pair with the sound source image to express its sound, we expect that the auxiliary use of the sound source image will guide the audio synthesis.

In this paper, we propose \textit{visual onoma-to-wave}, environmental sound synthesis from visual onomatopoeias.
Our method synthesizes sounds from onomatopoeia visual-text rather than onomatopoeia text. 
This enables us to control the detailed structure of the synthesized sound by transferring the visual concept of the visual onomatopoeia.
In other words, we introduce image stretching of visual onomatopoeias to control the duration of the environmental sound.
Furthermore, we introduce to condition the sound synthesis model by sound source images for controlling the overall impression of the sound.
This image-based auxiliary control makes it possible to synthesize more diverse sounds than label-based control with the conventional onoma-to-wave.
We also propose a data augmentation method to enhance the performance of environmental sound synthesis.
We conducted experimental evaluations to investigate 1) whether our methods appropriately transfer visual concepts to the sound and 2) whether they synthesizes diverse and natural sounds. The source code and demo are open-sourced on our project pages\footnote{\scriptsize{\url{https://sarulab-speech.github.io/demo_visual-onoma-to-wave/}}}. 
The contributions of this study are as follows:
    \vspace{-1.5mm}
    \begin{itemize} \leftskip -5.5mm \itemsep -0.5mm
        \item We propose visual onoma-to-wave, a new task and method of synthesizing environmental sounds from visual representations of onomatopoeias and sound sources.
        \item The proposed method and data augmentation allow synthesizing a more diverse sound with the same or better naturalness than the conventional method.
    \end{itemize}
    \vspace{-1mm}

\vspace{-3mm}
\section{Comparing sound synthesis methods} \vspace{-2mm} \label{sec:data_collection}
    \subsection{Sound synthesis from discrete symbols} \vspace{-2mm}
    The basic methods for environmental sound synthesis involve synthesizing from discrete labels representing sound events, e.g., ``cup'' or ``drum.''~\cite{kong2019sampleRNN,liu2021csg}
    Such methods use sound event labels to represent the overall impression of the synthesized sounds.
    The label is suitable in representing the overall impression of the synthesized sound but has no information to represent the detailed structure of the sound. 
    In contrast, text is a good means to describe the detailed structure.
    There are two types of methods for text-based control: describing all attributes of the sound in a sentence and separately describing the overall impression and detailed structure.
    Yang's method~\cite{yang2022diffsound} is of the former, and onoma-to-wave~\cite{okamoto2022Onoma}, 
    is of the latter.
    Since onomatopoeias efficiently imitate temporal changes of the sound, they are considered straightforward to control the detailed structure of the sound~\cite{lemaitre2014onoeffect1,sundaram2006onoeffect2}. 
    Onoma-to-wave also uses labels as auxiliary information in order to control the overall impression, following the label-based synthesis.
    
    \vspace{-2mm}
    \subsection{Sound synthesis from images} \vspace{-2mm}
    Methods for speech and sound synthesis from images have been proposed.
    Image-to-audio converts visual concepts in given images to synthesized sound~\cite{owens2016visually,zhou2018VtoA}.
    This image-based control of the overall expression of the sound is better in sound diversity than the aforementioned label-based control.
    In text-to-speech, visual text-to-speech (vTTS) replaces text of the standard text-to-speech with visual text (i.e., text as an image) and converts visual concepts in given visual text to synthesized speech~\cite{nakano2022vTTS}.
    This image-based control of the detailed structure of the speech enables a variety of speech expressions that are not possible with standard text-to-speech.
    
    In light of these methods, we introduce image representation into onoma-to-wave to enable the synthesis of diverse environmental sounds.
    Instead of onomatopoeia text, onomatopoeia visual-text (visual onomatopoeia) is used to synthesize.
    Not only textual but also visual concepts determine the detailed structure of the sound.
    Also, instead of a sound event label, a sound event image is used as an auxiliary to enhance the diversity in overall expression.
    Considering the application to comics and VR described in Section~\ref{sec:introduction}, the combination of visual onomatopoeia and auxiliary images is appropriate.
    Usually, a visual onomatopoeia corresponds to the source image in comics and VR.
    For example, in a baseball comic, the image of a baseball bat and the visual onomatopoeia of hitting a baseball are adjacent to each other. The bat image and visual onomatopoeia give the impression of audio to the reader.

\vspace{-3mm}
\section{Proposed methods} \vspace{-2mm}\label{sec:prop}
    \subsection{Visual onoma-to-wave} \label{sec:prop3.1} \vspace{-2mm}
        \textbf{Basic architecture.}
        Our method uses the Fastspeech2-inspired architecture~\cite{ren2020FS2,nakano2022vTTS} to learn models from visual onomatopoeias and sound event images as shown in \Fig{fig/prop_architecture}. 
        First, the visual onomatopoeia is sliced into individual characters. 
        As with vTTS~\cite{nakano2022vTTS}, we use visual onomatopoeias artificially generated from onomatopoeia texts.
        Assuming a monospaced font of the visual text, we slice the visual onomatopoeia into $n$ images of $h$-by-$w$ size, where $n, h, w$ are the number of characters and the height and width of each visual character, respectively. 
        These are a bit far from the realistic settings but sufficient to investigate the ideal performance of sound synthesis from visual onomatopoeias. 
        The sliced images are fed to the visual feature extractor, followed by the FastSpeech2-inspired encoder, variance adaptor\footnote{Since environmental sounds have no pitch-related feature, our variance adaptor does not have a pitch predictor that is used in the original FastSpeech2 in text-to-speech.}, and decoder. 
        The sound waveform is synthesized from the generated mel-spectrogram, using a neural vocoder~\cite{kong2020hifi}. 
        \\\textbf{Conditioning by sound event image.}
        A sound event image is used for controlling the overall expression of the sound. 
        The pre-trained CLIP~\cite{radford2021clip} image encoder encodes the input image to the sound event feature. 
        The obtained feature is added to the output of the variance adaptor. The CLIP pre-training involves a text-image contrastive objective; the pre-trained encoder is expected to extract meaningful features from an input image\footnote{We used the pre-trained CLIP image encoder to control the sound. Another control option is using the pre-trained CLIP text encoder, i.e., controlling the sound by natural language.}. 
        For example, a feature extracted from an image of a metal baseball bat corresponds to the text ``a metal baseball bat.'' This allows us to synthesize more diverse sounds than the sound event label, e.g., ``a baseball bat.''
        
        \drawfig{t}{0.98\linewidth}{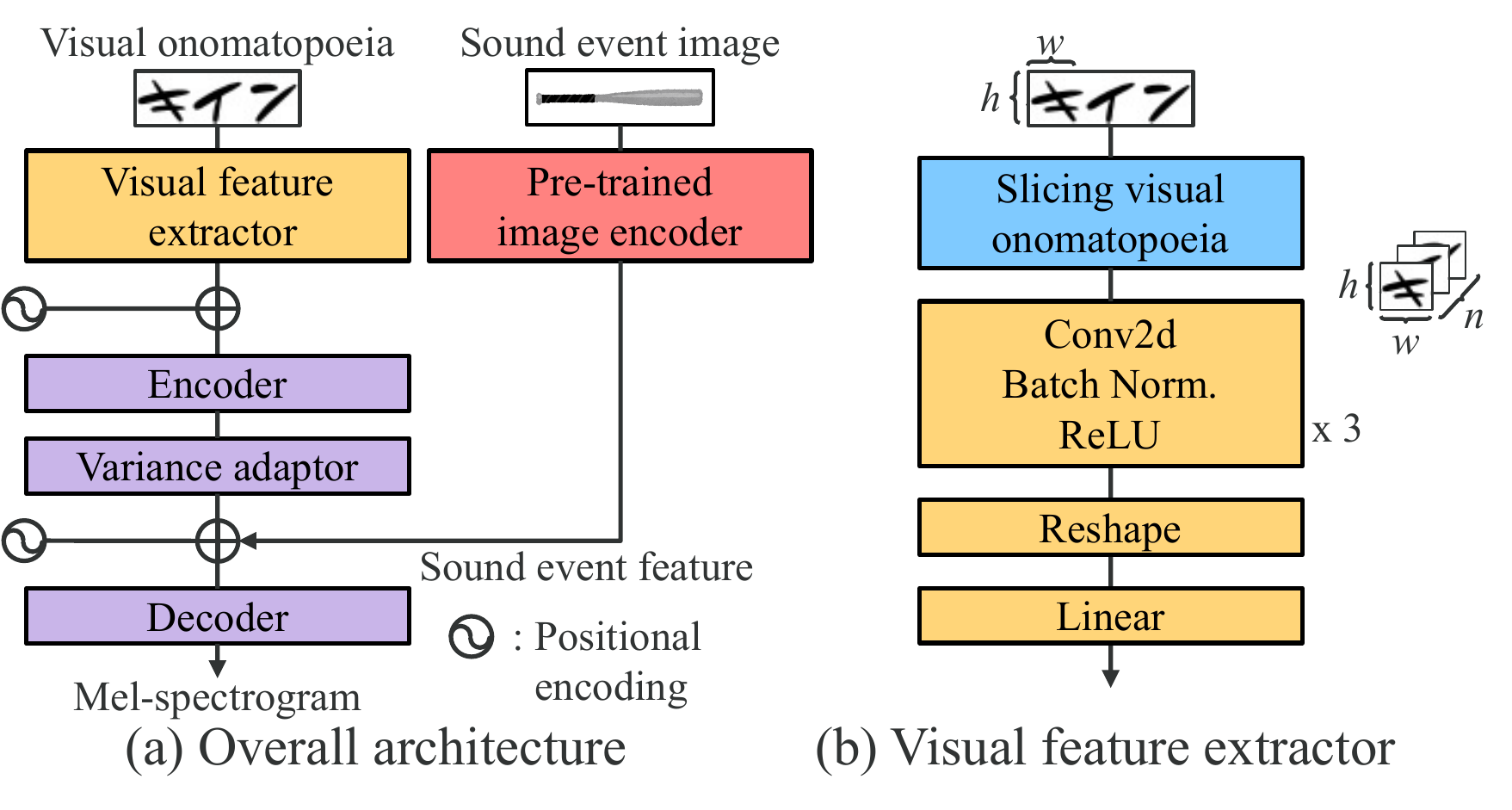}{Overall architecture of the proposed visual onoma-to-wave.} \label{fig:prop_architecture}
        
        \vspace{-2mm}
        \subsection{Transferring visual effects in visual onomatopoeia to sound} \vspace{-2mm}
        We introduce duration-informed stretch of a visual onomatopoeia for transferring its visual concept to sound. 
        As explained in Section~\ref{sec:introduction}, shrinking and expanding a visual onomatopoeia changes the duration of the imagined sound. 
        
        \drawfig{t}{0.9\linewidth}{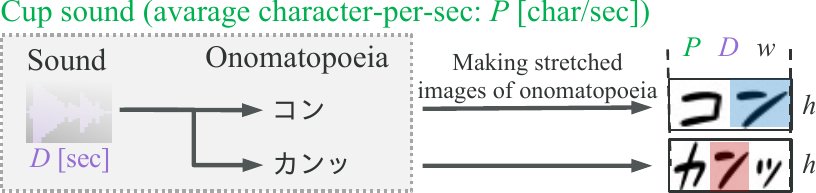}{Duration-informed stretch of visual onomatopoeia.\vspace{-2mm}} \label{fig:imgStretch}
        
        We first calculate the average sounding rate $P$~[character/sec] in each sound event cluster (sound event category in our study).
        This value indicates the number of onomatopoeia characters used for describing $1$-second sound. A width of a visual onomatopoeia corresponding to $D$-second sound is stretched at a rate of $P \cdot D$. 
        \Fig{fig/takamichi/imgStretch} shows an example of two different onomatopoeias corresponding to one sound. The onomatopoeias are visualized with the same shape, regardless of their number of characters.
        This method is applied to the training data, and there is no change in training. During inference, sound duration can be controlled by shrinking or expanding the visual onomatopoeia.
    
    \vspace{-2mm}    
    \subsection{Data augmentation} \vspace{-2mm}
    Frequent appearance of repetition is a unique characteristic of onomatopoeias~\cite{nasu2007ono_repeat}, e.g., ``ding-ding-ding-...'' (bell ringing repeatedly) and ``zzzzzz...'' (continued snoring sound). 
    We categorize repetition types into word- (e.g., ``ding-ding...'') and character-level (e.g., ``zzzz...''). 
    Our proposed data augmentation method synthesizes sounds from visual onomatopoeias\footnote{This can be also applied to the conventional onoma-to-wave.} with repetitions.
    
    \drawfig{t}{0.98\linewidth}{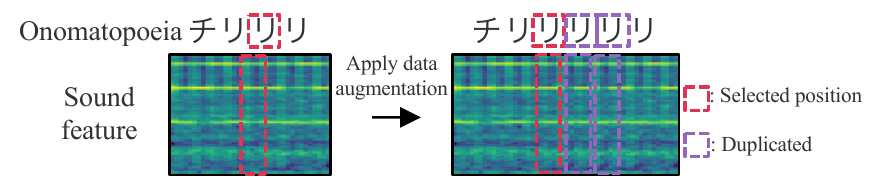}{Example of our data augmentation for character repetition.\vspace{-2mm}} 
    
    For word-level repetitions, we simply duplicate and concatenate a sound waveform corresponding to non-repetition onomatopoeias. 
    For example, the ``ding-ding'' sound is made by concatenating the ``ding'' sound. 
    For character-level repetition, we first find repeating ($\geq$ $3$ times) characters in the training data and temporally align characters and sound waveform. 
    We chose the middle part of the continuation, which is assumed more stable than the other parts, and duplicate the waveform segment at the subsequent position as shown in \Fig{fig/eg_charaug}. 
    This causes discontinuous sounding, but this may be alleviated by model training.

\vspace{-3mm}
\section{Experimental evaluation} \vspace{-2mm}\label{sec:experiment}
    \subsection{Experimental Setup} \label{sec:setup} \vspace{-2mm}
        \textbf{Dataset.}
        We used pairs of sounds and onomatopoeia texts of $10$ different sound events from RWCP-SSD~\cite{nakamura1999RWCP} (environmental sound corpus) and RWCP-SSD-Onomatopoeia~\cite{okamoto2020RWCPO} (onomatopoeia corpus aligning to RWCP-SSD), and used
        $13,170$ and $664$ audio samples for training and evaluation, respectively. 
        We filtered out onomatopoeias with confidence scores~\cite{okamoto2020RWCPO} lower than $3$ on a $5$-point scale. 
        Alignment between onomatopoeias and sound waveforms was obtained by training hidden Markov models using HTK~\cite{htk}. 
        We confirmed they were accurately aligned.
        \\\textbf{Visual onomatopoeias.}
        We used the ahaha-mojimoji\footnote{\scriptsize{\url{https://www.flopdesign.com/font7/ahahamojimoji-font}}} font and the Pillow module of python to artificially generate visual onomatopoeia from the onomatopoeia text.
        We used $24$-pt font, so $h=w=24$ when no image stretching was used.
        When using image stretching, the stretched image was adjusted to the same $w$ across all data using zero-padding.
        \\\textbf{Model and learning configuration.}
        The learning rate was set to $0.001$.
        In Sections \ref{sec:base}--\ref{sec:imgstr}, we used sound event labels as the auxiliary information rather than images. We used the same method as for speaker embedding~\cite{chien2021spEmb} to obtain 256-dimensional sound event features from the sound event labels.
        We used images as the auxiliary information in Section \ref{sec:soundimage}.
        The sound event feature was added to the output of the variance adaptor via a linear layer.
        Other configurations, e.g., model configuration and training schedule, followed the open-sourced implementation of vTTS~\cite{nakano2022vTTS}\footnote{\scriptsize{\url{https://github.com/Yoshifumi-Nakano/visual-text-to-speech}}}. 
        Original onoma-to-wave paper~\cite{okamoto2022Onoma} uses a Tacotron-inspired synthesis model~\cite{wang2017tacotron}, but in this paper, we used FastSpeech2-inspired model for both conventional onoma-to-wave and the proposed visual onoma-to-wave for fair comparison.

    \vspace{-2mm}
    \subsection{Evaluations} \vspace{-2mm}
    In \Sec{base}, we investigate whether visual onoma-to-wave is comparable in the basic performance to the conventional onoma-to-wave. In \Sec{imgstr}--\Sec{soundimage} we discuss our data augmentation method, image stretch-based control, and image-conditioned synthesis, respectively.
    
    Following the onoma-to-wave paper~\cite{okamoto2022Onoma}, we conducted five-scale mean opinion score (MOS) evaluations on three criteria: acceptance of synthetic sounds relative to onomatopoeias, expressiveness of the sounds relative to onomatopoeia, and naturalness of the sounds.
    Native Japanese speakers recruited through crowdsourcing participated in this evaluation. At least $20$ listeners participated in each MOS test.
        
        \vspace{-2mm}
        \subsubsection{Text input vs. Visual-text input} \label{sec:base} \vspace{-2mm}
        To evaluate the basic quality of the visual-text input, we compared synthetic sound samples from text (i.e., onoma-to-wave) or visual text (i.e., visual onoma-to-wave). 
        We used $664$ samples for each synthesis method.
        
        \begin{table}[t]
    \caption{MOS test results (mean$\pm$std) comparing text input and visual-text input. ``Reconstructed'' indicates reconstructed sound from ground-truth mel-spectrograms by neural vocoder.}
    \footnotesize
    \centering
    \begin{tabular}{r|ccc}
                                & Acceptance     & Expressiveness & Naturalness \\ \hline
        Reconstructed                & $3.50 \pm 1.14$ & $3.25 \pm 1.25$ & $3.40 \pm 1.15$ \\
        Onomatopoeia            & $3.61 \pm 1.21$ & $3.32 \pm 1.08$ & $3.21 \pm 1.22$ \\
        Visual onomatopoeia     & $3.63 \pm 1.12$ & $3.42 \pm 1.17$ & $3.27 \pm 1.22$
    \end{tabular}
    \vspace{-3mm}
    \label{tab:onoma-vs-visual-onoma}
\end{table}
        
        \Table{onoma-vs-visual-onoma} lists the results. 
        Since there is no statistical significance between text input and visual-text input for all criteria, we can say that visual onoma-to-wave is comparable to the conventional onoma-to-wave in basic performance.
        
        \vspace{-2mm}
        \subsubsection{Impact of data augmentation} \label{sec:aug} \vspace{-2mm}
        We then evaluated the impact of the proposed data augmentation. 
        For word-level repetition, the number of repetitions was set from $1$ through $2$ e.g., /dong/ $\rightarrow$ /dongdong/ (1) or /dongdongdong/ (2), in training and $0$ through $4$ for evaluation. Short words ($\leq 7$ characters) were augmented. Namely, the evaluation data include sounds repeated more than in the training data. For character-level repetition, the number of repetitions was set from $1$ through $5$, e.g., /dong/ $\rightarrow$ /doong/ (1) or /doooooong/ (5), in training and $0$ through $10$ for evaluation. We chose $11$ characters to be augmented, which covers $90$\% of repeated characters in the corpus. We compared synthesized sounds with and without data augmentation. Our main target is to evaluate the impact of data augmentation on visual onoma-to-wave, but we also applied data augmentation to the conventional onoma-to-wave. We used $1,000$ and $2,000$ audio samples for evaluation of word-level and character-level augmentation, respectively.

        For objective evaluation, we calculated sound duration along with the word or character repetitions. The duration of non-repeated onomatopoeia was normalized to $1.0$, and the relative changes in duration was calculated. For subjective evaluation, we conducted MOS tests as in Section \ref{sec:setup}.

        \drawfig{t}{0.98\linewidth}{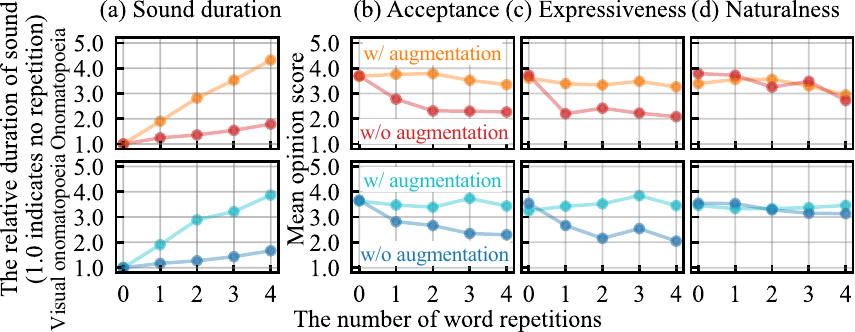}{Results of objective and subjective evaluation for data augmentation of word-level repetition.}
        
        \drawfig{t}{0.98\linewidth}{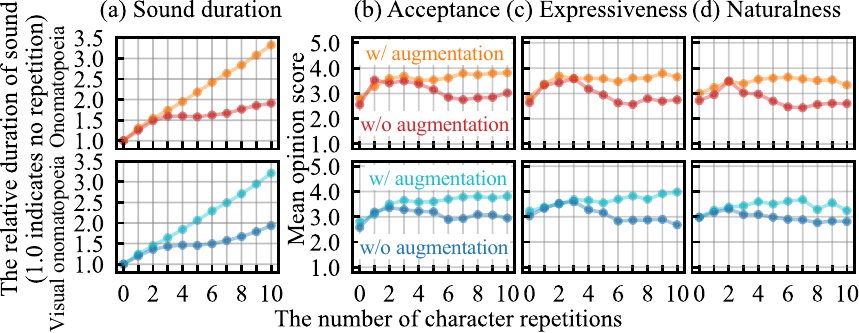}{Results of objective and subjective evaluation for data augmentation of character-level repetition.}
        \Fig{fig/takamichi/422word_plot}~(a) and \Fig{fig/takamichi/422char_plot}~(a) show the results of the objective evaluation. Training without data augmentation did not increase the duration along with the number of repetitions. Since FastSpeech2 explicitly assigns duration to each character, the repeated words and characters had non-zero durations. However, we found that the durations of the repeated ones were very close to zero; increasing repetitions no longer change the total duration. In contrast, our data augmentation method solves this problem, and the duration changes near-linearly. 

        \Fig{fig/takamichi/422word_plot}~(b)--(d) and \Fig{fig/takamichi/422char_plot}~(b)--(d) show the results of the subjective evaluation. 
        As shown in \Fig{fig/takamichi/422word_plot}~(d), our data augmentation method did not lose the MOS in naturalness compared with no data augmentation. 
        In contrast, from \Fig{fig/takamichi/422word_plot}~(b)--(d), our data augmentation method had better MOSs in acceptance and expressiveness than no augmentation. 
        As mentioned above, synthesized sounds without data augmentation do not respond to the increase in repetitions. This causes perceptual mismatch between the onomatopoeia and synthetic sound. 
        Our data augmentation method efficiently solves this and preserves MOSs even if synthesizing from onomatopoeias with many repetitions. 
        The same tendency was observed in character-level repetitions, as shown in \Fig{fig/takamichi/422char_plot}~(b)--(d), but significant improvements with our data augmentation method were observed for naturalness. 
        
        \vspace{-2mm}
        \subsubsection{Duration control by image stretching} \label{sec:imgstr} \vspace{-2mm}
        We evaluated duration control by image stretching.
        The stretch ratios were set to ${0.5, 1.0, 1.5, 2.0}$, and $200$ audio samples were used for each stretch ratio. 
        For comparison, we trained models without pre-processing of image stretching. 
        The objective evaluation was on the relative duration of the sound, setting the duration without stretching (i.e., the ratio of $1.0$) to $1.0$. 
        The subjective evaluation was conducted in the same manner as in Section \ref{sec:aug}.
        
        \drawfig{t}{0.98\linewidth}{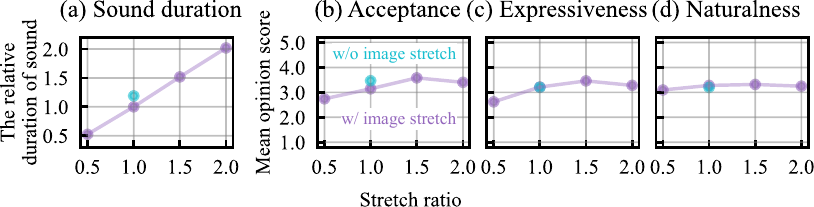}{Results of objective and subjective evaluations for image stretching-based duration control.}
        
        \Fig{fig/takamichi/423_plot} shows the results. From \Fig{fig/takamichi/423_plot}~(a), the sound duration changed linearly, so we can say that stretching-based duration control performs well. 
        In contrast, as shown in \Fig{fig/takamichi/423_plot}~(b)--(d), while the MOS on naturalness were maintained among the ratios, those on acceptance and expressiveness degraded when the ratio was $0.5$. We will investigate the reason for this in the future.
        
        \vspace{-2mm}
        \subsubsection{Synthesis of diverse sounds from sound event image} \label{sec:soundimage} \vspace{-2mm}
        We evaluated the quality and diversity of sound synthesized from visual onomatopoeias and auxiliary images. 
        We prepared additional sound-onomatopoeia-image pairs for the evaluation. 
        For sound events in RWCP-SSD, we crawled the Internet and collected $50$ photos per event, which matched the sounds. 
        Additionally, we prepared a new sound event class ``baseball bat'' to evaluate the synthesized sound. 
        We crawled the Internet to collect $150$ photos of baseball bats made of metal, wood, and plastic. 
        GameSynth\footnote{\scriptsize{\url{http://tsugi-studio.com/web/en/products-gamesynth.html}}}, a sound effect creation tool, was used to artificially generate the corresponding sounds. 
        The corresponding onomatopoeias were made from crowdsourcing as in the onoma-to-wave paper~\cite{okamoto2022Onoma}, and at least eight onomatopoeias were obtained for each sound. $267$ sound-onomatopoeia-image pairs were used for evaluation, and the remaining were for the training data. 
        The RWCP-SSD sound event data was used only for training and not for evaluation.
        There was no overlap in sounds, onomatopoeias, and images between the training and evaluation data. 
        Using these data, we trained the photo-conditioned synthesis model and evaluated the synthetic sounds. 
        For application to comics, we also trained the line-drawing-conditioned synthesis model. 
        Anime2Sketch~\cite{xiang2022Sketch2anime}\footnote{\scriptsize{\url{https://github.com/Mukosame/Anime2Sketch}}} was used to convert the obtained photos to the line-drawing style. 
        For comparison, we also trained a label-conditioned model, i.e., the ``baseball bat'' label was used for synthesis, the same as with the conventional onoma-to-wave. 
        We used ViT-L/14~\cite{dosovitskiy2020ViT}, as the CLIP~\cite{radford2021clip} image encoder\footnote{\scriptsize{\url{https://github.com/openai/CLIP}}}. 
        The objective evaluation involved sound diversity as a mean squared distance between all combinations of the synthesized sounds~\cite{tamaru2020generative}. 
        The subjective evaluation was conducted in the same manner as in Section~\ref{sec:imgstr}.

        \drawfig{t}{0.8\linewidth}{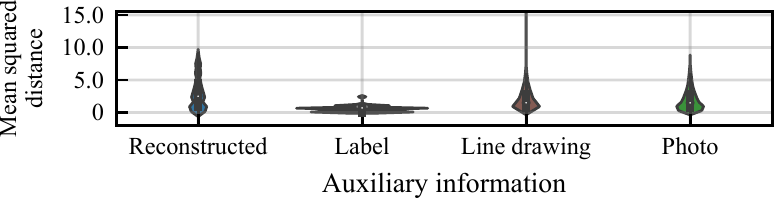}{Sound diversity in synthesis from sound event image.\vspace{-2mm}}
        
        \begin{table}[t]
    \caption{MOSs comparing auxiliary information. ``Reconstructed'' indicates reconstructed sound from ground-truth mel-spectrograms by neural vocoder. \textbf{Bold} means that there is significant difference between photo- or line-drawing-conditioned and label-conditioned.}
    \footnotesize
    \centering
    \begin{tabular}{r|ccc}
                                & Acceptance     & Expressiveness & Naturalness \\ \hline
        Reconstructed                & $3.60 \pm 1.13$ & $3.43 \pm 1.16$ & $3.32 \pm 1.14$ \\
        Line drawing            & $\mathbf{2.86 \pm 1.25}$ & $\mathbf{2.97 \pm 1.12}$ & $3.46 \pm 0.90$ \\
        Photo     & $\mathbf{2.87 \pm 1.35}$ & $\mathbf{2.90 \pm 1.11}$ & $3.31 \pm 0.97$ \\
        Label & $2.49 \pm 1.11$ & $2.50 \pm 1.08$ & $\mathbf{3.70 \pm 0.91}$
    \end{tabular}
    \vspace{-5mm}
    \label{tab:soundimage}
\end{table}
        
        \Fig{fig/takamichi/424violin_plot} and \Table{soundimage} show the results. From \Fig{fig/takamichi/424violin_plot}, the distributions of sounds of photo-conditioned and line-drawing-conditioned synthesis were close to that of the reconstructed sounds. 
        On the other hand, label-conditioned synthesis was less in diversity compared with the others. 
        We confirmed that the use of sound event images enables us to synthesize more diverse sounds than using labels. 
        From \Table{soundimage}, the photo- and line-drawing-conditioned synthesis achieved significantly higher MOSs on acceptance and expressiveness than the label-conditioned synthesis. 
        Therefore, we can say that sound event images help synthesize appropriate sounds. 
        In contrast, we also observed that it degrades the naturalness MOSs.
        This may be because label-conditioned synthesis uniquely determines features, whereas photo- and line-drawing-conditioned synthesis do not, thus making the synthesis slightly unstable.
        We will address this for future work.

\vspace{-2mm}
\section{Conclusion} \vspace{-2mm}\label{sec:conclusion}
    We proposed visual onoma-to-wave, which uses image representations to synthesize environmental sounds from onomatopoeias.
    For future work, we plan to synthesize sound from real data.

\printbibliography

@inproceedings{lloyd2011sound,
  title={Sound synthesis for impact sounds in video games},
  author={Lloyd, D Brandon and Raghuvanshi, Nikunj and Govindaraju, Naga K},
  booktitle={Symposium on Interactive 3D Graphics and Games},
  pages={55--62},
  year={2011}
}

@article{okamoto2022Onoma,
  title={Onoma-to-wave: Environmental sound synthesis from onomatopoeic words},
  author={Okamoto, Yuki and Imoto, Keisuke and Takamichi, Shinnosuke and Yamanishi, Ryosuke and Fukumori, Takahiro and Yamashita, Yoichi and others},
  journal={APSIPA Transactions on Signal and Information Processing},
  volume={11},
  number={1},
  year={2022},
  publisher={Now Publishers, Inc.}
}

@inproceedings{baek2022COO,
  title={{COO}: Comic Onomatopoeia Dataset for Recognizing Arbitrary or Truncated Texts},
  author={Baek, Jeonghun and Matsui, Yusuke and Aizawa, Kiyoharu},
  booktitle={Proceedings of the European Conference on Computer Vision (ECCV)},
  year={2022}
}

@inproceedings{nakano2022vTTS,
  title={{vTTS}: visual-text to speech},
  author={Nakano, Yoshifumi and Saeki, Takaaki and Takamichi, Shinnosuke and Sudoh, Katsuhito and Saruwatari, Hiroshi},
  booktitle={Proceedings of the IEEE Spoken Language Technology Workshop (SLT)},
  year={2022}
}

@inproceedings{zhou2018VtoA,
  title={Visual to sound: Generating natural sound for videos in the wild},
  author={Zhou, Yipin and Wang, Zhaowen and Fang, Chen and Bui, Trung and Berg, Tamara L},
  booktitle={Proceedings of the IEEE Conference on Computer Vision and Pattern Recognition},
  pages={3550--3558},
  year={2018}
}

@article{
anony2023audiogen,
title={{AudioGen}: Textually Guided Audio Generation},
author={Felix Kreuk and Gabriel Synnaeve and Adam Polyak and Uriel Singer and Alexandre D\'{e}fossez and Jade Copet and Devi Parikh and Yaniv Taigman and Yossi Adi},
journal={arXiv preprint arXiv:2209.15352},
year={2022}
}

@article{wang2019comic,
  title={Comic-guided speech synthesis},
  author={Wang, Yujia and Wang, Wenguan and Liang, Wei and Yu, Lap-Fai},
  journal={ACM Transactions on Graphics (TOG)},
  volume={38},
  number={6},
  pages={1--14},
  year={2019},
  publisher={ACM New York, NY, USA}
}

@inproceedings{owens2016visually,
  title={Visually indicated sounds},
  author={Owens, Andrew and Isola, Phillip and McDermott, Josh and Torralba, Antonio and Adelson, Edward H and Freeman, William T},
  booktitle={Proceedings of the IEEE Conference on Computer Vision and Pattern Recognition},
  pages={2405--2413},
  year={2016}
}

@inproceedings{kong2019sampleRNN,
  title={Acoustic scene generation with conditional {SampleRNN}},
  author={Kong, Qiuqiang and Xu, Yong and Iqbal, Turab and Cao, Yin and Wang, Wenwu and Plumbley, Mark D},
  booktitle={Proceedings of IEEE International Conference on Acoustics, Speech and Signal Processing (ICASSP)},
  pages={925--929},
  year={2019}
}

@INPROCEEDINGS{liu2021csg,
  author={Liu, Xubo and Iqbal, Turab and Zhao, Jinzheng and Huang, Qiushi and Plumbley, Mark D. and Wang, Wenwu},
  booktitle={2021 IEEE 31st International Workshop on Machine Learning for Signal Processing (MLSP)}, 
  title={Conditional Sound Generation Using Neural Discrete Time-Frequency Representation Learning}, 
  year={2021},
  volume={},
  number={},
  pages={1-6},
  doi={10.1109/MLSP52302.2021.9596430}}

@article{yang2022diffsound,
  title={Diffsound: Discrete Diffusion Model for Text-to-sound Generation},
  author={Yang, Dongchao and Yu, Jianwei and Wang, Helin and Wang, Wen and Weng, Chao and Zou, Yuexian and Yu, Dong},
  journal={arXiv preprint arXiv:2207.09983},
  year={2022}
}

@inproceedings{oh2019VReffect,
  title={The effect of onomatopoeia to enhancing user experience in virtual reality},
  author={Oh, Jiwon and Kim, Gerard J},
  booktitle={Proceedings of International Conference on Human-Computer Interaction},
  pages={143--152},
  year={2019},
  organization={Springer}
}

@article{lemaitre2014onoeffect1,
  title={On the effectiveness of vocal imitations and verbal descriptions of sounds},
  author={Lemaitre, Guillaume and Rocchesso, Davide},
  journal={The Journal of the Acoustical Society of America},
  volume={135},
  number={2},
  pages={862--873},
  year={2014},
  publisher={Acoustical Society of America}
}

@inproceedings{sundaram2006onoeffect2,
  title={Vector-based Representation and Clustering of Audio Using Onomatopoeia Words},
  author={Sundaram, Shiva and Narayanan, Shrikanth S},
  booktitle={AAAI Fall Symposium: Aurally Informed Performance},
  pages={55},
  year={2006}
}

@inproceedings{chien2021spEmb,
  title={Investigating on incorporating pretrained and learnable speaker representations for multi-speaker multi-style text-to-speech},
  author={Chien, Chung-Ming and Lin, Jheng-Hao and Huang, Chien-yu and Hsu, Po-chun and Lee, Hung-yi},
  booktitle={Proceedings of IEEE International Conference on Acoustics, Speech and Signal Processing (ICASSP)},
  pages={8588--8592},
  year={2021},
  organization={IEEE}
}

@article{ren2020FS2,
  title={Fastspeech 2: Fast and high-quality end-to-end text to speech},
  author={Ren, Yi and Hu, Chenxu and Tan, Xu and Qin, Tao and Zhao, Sheng and Zhao, Zhou and Liu, Tie-Yan},
  journal={arXiv preprint arXiv:2006.04558},
  year={2020}
}

@inproceedings{okamoto2020RWCPO,
  title={{RWCP-SSD-Onomatopoeia}: Onomatopoeic Word Dataset for Environmental Sound Synthesis},
  author={Okamoto, Yuki and Imoto, Keisuke and Takamichi, Shinnosuke and Yamanishi, Ryosuke and Fukumori, Takahiro and Yamashita, Yoichi},
  booktitle={Proceedings of Detection and Classification of Acoustic Scenes and Events (DCASE)},
  pages={125--129},
  year={2020}
}

@article{nakamura1999RWCP,
  title={Sound scene data collection in real acoustical environments},
  author={Nakamura, Satoshi and Hiyane, Kazuo and Asano, Futoshi and Endo, Takashi},
  journal={Journal of the Acoustical Society of Japan (E)},
  volume={20},
  number={3},
  pages={225--231},
  year={1999},
  publisher={Acoustical Society of Japan}
}

@article{kong2020hifi,
  title={Hifi-gan: Generative adversarial networks for efficient and high fidelity speech synthesis},
  author={Kong, Jungil and Kim, Jaehyeon and Bae, Jaekyoung},
  journal={Advances in Neural Information Processing Systems},
  volume={33},
  pages={17022--17033},
  year={2020}
}

@inproceedings{radford2021clip,
  title={Learning transferable visual models from natural language supervision},
  author={Radford, Alec and Kim, Jong Wook and Hallacy, Chris and Ramesh, Aditya and Goh, Gabriel and Agarwal, Sandhini and Sastry, Girish and Askell, Amanda and Mishkin, Pamela and Clark, Jack and others},
  booktitle={Proceedings of International Conference on Machine Learning},
  pages={8748--8763},
  year={2021},
  organization={PMLR}
}

@article{dosovitskiy2020ViT,
  title={An image is worth 16x16 words: Transformers for image recognition at scale},
  author={Dosovitskiy, Alexey and Beyer, Lucas and Kolesnikov, Alexander and Weissenborn, Dirk and Zhai, Xiaohua and Unterthiner, Thomas and Dehghani, Mostafa and Minderer, Matthias and Heigold, Georg and Gelly, Sylvain and others},
  journal={arXiv preprint arXiv:2010.11929},
  year={2020}
}

@article{nasu2007ono_repeat,
  title={The word-final moraic obstruent in {Japanese} mimetics},
  author={Nasu, Akio},
  journal={Journal of the Phonetic Society of Japan},
  volume={11},
  number={1},
  pages={47--57},
  year={2007},
  publisher={Phonetic Society of Japan}
}

@inproceedings{xiang2022Sketch2anime,
  title={Adversarial Open Domain Adaptation for Sketch-to-Photo Synthesis},
  author={Xiang, Xiaoyu and Liu, Ding and Yang, Xiao and Zhu, Yiheng and Shen, Xiaohui and Allebach, Jan P},
  booktitle={Proceedings of the IEEE/CVF Winter Conference on Applications of Computer Vision},
  pages={1434--1444},
  year={2022}
}

@article{wang2017tacotron,
  title={Tacotron: Towards end-to-end speech synthesis},
  author={Wang, Yuxuan and Skerry-Ryan, RJ and Stanton, Daisy and Wu, Yonghui and Weiss, Ron J. and Jaitly, Navdeep and Yang, Zongheng and Xiao, Ying and Chen, Zhifeng and Bengio, Samy and others},
  journal={arXiv preprint 1703.10135},
  year={2017}
}

@article{huang2014fluidity,
  title={Fluidity of modes in the translation of manga: The case of {Kishimoto’s} Naruto},
  author={Huang, Cheng-Wen and Archer, Arlene},
  journal={Journal of Visual Communication},
  volume={13},
  number={4},
  pages={471--486},
  year={2014},
  publisher={Sage Publications Sage UK: London, England}
}

@article{tamaru2020generative,
  title={Generative moment matching network-based neural double-tracking for synthesized and natural singing voices},
  author={Tamaru, Hiroki and Saito, Yuki and Takamichi, Shinnosuke and Koriyama, Tomoki and Saruwatari, Hiroshi},
  journal={IEICE Transactions on Information and Systems},
  volume={103},
  number={3},
  pages={639--647},
  year={2020},
  publisher={The Institute of Electronics, Information and Communication Engineers}
}

@misc{htk,
  title={Hidden Markov Model Toolkit ({HTK})},
  howpublished={\url{https://htk.eng.cam.ac.uk/}}

}
\end{document}